\documentclass[aps,prd,twocolumn,showpacs,preprintnumbers,amsmath,amssymb,nofootinbib]{revtex4}
\usepackage{bm}
\usepackage{amsmath}
\renewcommand{\onlinecite}[1]{\citep{#1}}

\begin{document}
\hyphenation{Max-well Lor-entz}
\preprint{hep-th/0305084}
\title{Spin-1/2 Maxwell Fields}
\author{Rollin S. Armour, Jr.}
\affiliation{Physics Department, Mercer University, Macon, GA 31207-0001, USA}%
\email{armour_r@mercer.edu}
\date{May 9, 2003; revised March 17, 2004}



\begin{abstract}
\hspace{.2cm}\small{Requiring covariance of Maxwell's equations without {\it a priori\/} imposing
charge invariance allows for both spin-1 and spin-1/2 transformations of the complete Maxwell
field and current. The spin-1/2 case yields new transformation rules, with new invariants, for all
traditional Maxwell field and source quantities. The accompanying spin-1/2 representations of the
Lorentz group employ the Minkowski metric, and consequently the primary spin-1/2 Maxwell
invariants are also spin-1 invariants; for example, $\Phi^2 - {\bf A}^2$, ${\bf E}^2 - {\bf B}^2 +
2i {\bf E} \bm{\cdot} {\bf B} - ({\partial}_{0}{\Phi} + {\bm{\nabla \cdot}}{\bf A})^2$. The
associated Maxwell Lagrangian density is also the same for both spin-1 and spin-1/2 fields.
However, in the spin-1/2 case, standard field and source quantities are complex and both charge
and gauge invariance are lost. Requiring the potentials to satisfy the Klein-Gordon equation
equates the Maxwell and field-potential equations with two Dirac equations of the Klein-Gordon
mass, and thus one complex Klein-Gordon Maxwell field describes either two real vector fields or
two Dirac fields, all of the same mass.}





\end{abstract}
\maketitle

\section{\label{sec:intro}INTRODUCTION}

The standard Lorentz transformation properties of the Maxwell field derive from more than simply
covariance of Maxwell's equations. The usual vector or tensor properties of the field include an
additional requirement, generally fixing the transformation rule of the source terms
\cite{Sachs1,Ros}. Traditionally, this requirement is filled by imposing invariance on electric
charge \cite{Ros,Jax}. Charge invariance forces the charge density to transform as the top
component of a four-vector, leading to the usual transformation properties for the remaining field
and source terms \cite{Ros}. But the additional constraint may take other forms. Among these are
explicit covariance of the source-free Maxwell equations (required by Einstein \cite{Einstein}),
imposing an invariant gauge-fixing condition $\Gamma = {\partial}_{0}{\Phi} + {\bm{\nabla
\cdot}}{\bf A}$ \cite{Aitch}, equating the charge-density current with the Dirac
probability-density current \cite{B&D}, and simultaneous charge invariance and covariance of the
Lorentz four-force \cite{Jax}. Each of these requirements in conjunction with Maxwell equation
covariance leads to the expected transformation properties of the Maxwell field. However, {\it a
priori\/} imposing such constraints on Lorentz transformations of Maxwell's equations is a
restricted case of the more general requirement of their covariance. In the general case, we may
allow the Maxwell variables to transform in any manner leaving the equations as a whole covariant,
and then determine the specific transformation properties of the various field and source terms.
Our interest here is in the possibility that the general case may allow alternate transform rules
for the Maxwell variables.

Indeed, there is some evidence in the literature to support this interest. Previous authors have
found that specialized or restricted forms of Maxwell's equations describe spin-1/2 fields
\cite{Darwin,Made,Lj,Sal,Sal1,Sal2,Sim0,Sim1,Sim2,Sim3,Bruce,Pap}, and in one instance a
four-vector combined electric and magnetic field \cite{Sim3}. Both observations are a result of
efforts to describe the Dirac field in electromagnetic (EM) terms
\cite{Lj,Sal,Sal1,Sal2,Sim0,Sim1,Sim2,Sim3,Bruce,Pap,Milner,Koga,Krug,8comp,Camp,RV,
RV1,RVR,Darwin,Made}. Darwin \cite{Darwin} alluded to the first possibility in 1928, not long
after the Dirac equation debuted, noting that it was equivalent to a massive form of Maxwell's
equations with gradient-constrained electric and magnetic currents
$j^\alpha_{elec}=\partial^\alpha f,~j^\alpha_{mag} =\partial^\alpha h$. A series of authors have
since rediscovered similar Maxwell-Dirac equation correlations
\cite{Lj,Sal,Sal1,Sal2,Sim0,Sim1,Sim2,Sim3,Pap,Bruce,Milner,Koga,Krug,8comp}, primarily
illustrating spin-1/2 Maxwell-like fields, but among these are Darwin-constrained Maxwell
equations with multiple transformation properties.

In one such approach, Sallhofer \cite{Sal,Sal1,Sal2,Sim0} describes a direct correspondence
between the Dirac equation and the source-free Maxwell equations in a medium of potential- and
mass-dependent electromagnetic permittivity and permeability ({\it a la\/} Madelung \cite{Made}).
The correspondence gives 1/2-angle factors to solutions of the source-free Maxwell equations
allowing for an electromagnetic-like description of a spin-1/2 matter wave. ``Electrodynamic
particle fields with half-integer spin are represented ... as a standing [EM] wave with a
cancelling counter-wave'' \cite{Sal1}, rather like a dipole, and thus do not appear
electromagnetic from the outside \cite{Sal2}. But the associated Maxwell-Dirac relationship is not
Lorentz covariant, leaving unaddressed the question of spin-1/2 transformation properties of the
Maxwell field itself.

However, in a Lagrangian study of the Darwin-Maxwell equations, potentials and magnetic current
included, Ljolje \cite{Lj,8comp} finds that ``The [massless] Dirac equation or equivalent
[massless Darwin-Maxwell] equations ... [each] describe two different physical fields: a bispinor
field, and a field composed of a second order tensor field and one scalar and one pseudoscalar
field'' \cite{Lj}. The associated Lagrangian density is invariant under {\it both\/} spin-1 and
spin-1/2 (Dirac spinor) transformations in both massless and massive cases. From the Maxwell
perspective, the formalism is restricted primarily by the Darwin gradient-constrained sources, but
Ljolje gives a thorough account of the equivalence of these special-source Maxwell equations and
the Dirac equation, illustrating a fermionic Maxwell-like field. Several related
investigations also describe spin-1/2, or Darwin-based, Maxwell-like fields (see especially Papini
\onlinecite{Pap}) \cite{Bruce,Milner,Koga,Krug}, while many others have explored similar Dirac- or
fermion-like Maxwell equations
\cite{guide,Rumer,Oppy,Maj,Im,Oh,Kobe,Moli,Arch,Good,Moses1,Moses2,Moses3,Sachs1,Leit,Vach,
Dow,Dow1,Jena,Chow,Weyl,BB,Abreu,Sipe,Drum,Dvoe1,Gian,Esp,Gerst,Gerst1,Ed,Mun,Geurdes,Hill,
daS,Jank,Dav,arm,Dvoe2,Dvoe3,Gsponer,Bade}, a number of which we will refer to in progress below
\cite{guide,clear}.

Simulik and Krivsky \cite{Sim1,Sim2,Sim3} have since solved the Lorentz covariance problem of the
Sallhofer model by including these same gradient-constrained electric and magnetic currents,
yielding the massless Darwin and Dirac equations. Mass appears as a property of the EM medium, as
before. Called the ``maximally symmetric'' Maxwell equations, a real Maxwell field is written with
four complex components $({\bm E}-i{\bm B}, f-ih) \equiv \psi_{MS}$, the gradient of the fourth
returns the electric and magnetic currents, giving the Darwin sources (no potentials included). In
the massless case, the authors have explicitly demonstrated covariance of this form of Maxwell's
equations while $\psi_{MS}$ transforms under the Dirac representation of the Lorentz group
$(1/2,0)\oplus(0,1/2)$, the expected tensor-scalar representation $(1,0)\oplus(0,0)$ \cite{Sim3},
and the vector representation $(1/2,1/2)$.\footnote{In a separate approach, Esposito
\onlinecite{Esp} and T. Ivezic [{\it Found. Phys.\/} {\bf 33}, 1339 (2003) (hep-th/0302188);
physics/0311043; and references therein] define ``auxiliary'' electric and magnetic four-vectors
from the electromagnetic field tensor, its dual, and the four-velocity, $E^\alpha =
F^{\alpha\beta}v_\beta/c$ and $B^\alpha = \widetilde{F}^{\alpha\beta}v_\beta/c$. This leads to
manifestly-covariant, though non-Maxwellian, definitions of electric and magnetic fields. But the
transformation properties of the field tensors, whose elements are the standard electric and
magnetic field components, are unchanged. Hence this formalism redescribes the traditional tensor
Maxwell field.} In fact, though unmentioned by the authors, this specialized form of Maxwell's
equations, like their equivalent massless Dirac equation, transforms covariantly under {\it any\/}
four-dimensional representation of the Lorentz group (see beginning of Sec. \ref{sec:trans}
below). However, the significance of its specialized sources is not clear (see Gersten, Refs.
\onlinecite{Sim1,Gerst}), obscuring comparison with the standard Maxwell field.

The importance of the above investigations for our purposes is not in their electromagnetic-like
descriptions of the Dirac field, which invariably require a constrained form of Maxwell's
equations \cite{Gsponer}, but in their demonstration of Maxwell-like fields with spin-1/2
transformation properties. Here, we will find that a more significant spin-1/2 Lorentz
transformation property is characteristic of the {\it complete Maxwell field without
constraints\/}.

We begin by revisiting the Maxwell and field-potential equations in a familiar matrix form,
obtained by factoring a simple 4-d wave equation. We then readdress the now century-old problem of
Maxwell-equation covariance \cite{Lorentz,Einstein}, allowing charge, the gauge-fixing condition,
and charge-density current to transform without additional requirements. Covariance of Maxwell's
equations then naturally returns {\it two\/} transformation rules for the full Maxwell field, the
expected spin-1 rule and a new spin-1/2 transformation of all Maxwell variables.

The resulting spin-1/2 Maxwell field differs significantly from its previously known vector and
spinor forms. First, it is the complete field and current, not a restricted or partial form.
Second, its spin-1/2 transformation employs the $(1/2,0)\oplus(1/2,0)$ and its conjugate
$(0,1/2)\oplus(0,1/2)$ representations of the Lorentz group, {\it not\/} the Dirac representation,
the first transforming the potentials and sources, and the second the fields. Surprisingly, the
natural basis for these two representations is {\it Minkowskian\/}, with group elements $R$
satisfying the orthogonality condition $R^TgR=g$ where $g$ is the {\it Minkowski\/} metric. That
is, the Minkowski metric accompanies the Maxwell invariance relations under both spin-1 and {\it
spin-1/2\/} transformations, serving as both four-vector and {\it four-spinor\/} metrics. As a
result, the primary spin-1/2 Maxwell invariants are also spin-1 invariants, e.g., $\Phi^2 - {\bf
A}^2$, ${\bf B}^{2} - 2{i\/}{\bf E}\bm{\cdot}{\bf B} - {\bf E}^{2} + (\partial_0 \Phi +
{\bm{\nabla \cdot}}{\bf A})^2$. However, spin-1/2 field and source terms are generally complex and
neither charge nor the four-divergence of the potentials remain invariant. Separating these terms
into real and imaginary parts gives standard real electric and magnetic charge-density currents
and a corresponding real Maxwell field in all frames. Third, like the field equations themselves,
the Lagrangian densities for the spin-1/2 and spin-1 Maxwell fields are {\it identical.\/} That
is, the Lagrangian density for the complete Maxwell equations in the matrix form employed here is
a scalar under {\it both spin-1 and spin-1/2\/} transformations of the Maxwell variables.
Introducing a Proca mass term in the Lagrangian density returns the Proca-Maxwell equations which
are {\it also\/} covariant under both spin-1 and spin-1/2 Lorentz transformations.

Finally, the correspondence between the Maxwell and Dirac equations takes a natural form, similar
to that of Ljolje \cite{Lj}. Requiring the EM potentials to satisfy the Klein-Gordon equation, as
if to describe a massive vector boson, equates the Maxwell and field-potential equations with two
Dirac equations of identical mass. The resulting Maxwell-Dirac equation correspondence is thus
one-to-two; or equivalently, one complex Klein-Gordon Maxwell field can be understood as either
two real vector fields or two complex Dirac fields.

Our work is arranged as follows: Sec. \ref{sec:MaxEq}, Matrix Maxwell Equations; \ref{sec:trans},
Spin-1/2 and Spin-1 Transformations of the Maxwell Variables; \ref{sec:four}, The Minkowski
Spinor; \ref{sec:invar}, Spin-1/2 Electromagnetic Invariants; \ref{sec:lagrangian}, Dual
Spin-1/2--Spin-1 Maxwell and Proca Lagrangian Densities; \ref{sec:halfEM}, Loss of Gauge
Invariance, Charge Invariance, and the Lorentz Four-Force in the Spin-1/2 Case; \ref{sec:MKGM},
Equivalence of the Klein-Gordon Maxwell and Two Dirac Equations; \ref{sec:concl}, Conclusion;
Appendices, \ref{sec:covar}: Covariance of Maxwell's Equations; \ref{sec:nota}: Index Notation;
\ref{sec:basis}: Spin-1/2 and Spin-1 Minkowski-Basis Eigenvectors.

\section{\label{sec:MaxEq}THE MATRIX MAXWELL EQUATIONS}

Expanding on the work of Moses \cite{Moses1,Moses2,Moses3}, we begin with\footnote{Greek indices
are 0-3, $x^{0}=ct$, $c=1$, $\text{tr}{\,}({g\/}_{{\alpha}{\beta}}) = -2$.}
\begin{equation}\label{AY}
\partial_\alpha \partial^\alpha {\rm A} = {\rm Y}
\end{equation}
and factor the d'Alembertian operator via
\begin{equation}\label{fact}
\partial_\alpha \partial^\alpha {\rm A} = (\mu^{\alpha} \partial^{\alpha}) (\mu^{\beta}
\partial_{\beta}) {\rm A} = {\rm Y}.
\end{equation}
Equation (\ref{fact}) separates into two first-order equations
\begin{subequations} \label{MME}
\begin{align}
\mu^{\alpha} \partial^{\alpha} {\rm A} &= {\rm C} \label{MMEa}
\\ \mu^{\beta} \partial_{\beta} {\rm C} &= {\rm Y}, \label{MMEb}
\end{align}
\end{subequations}
where A, C, and Y remain to be interpreted. The matrices $\mu^\alpha$ must satisfy $(\mu^\alpha)^2
= {\bf 1}$, $\mu^0 \mu^j = \mu^j \mu^0$, $\mu^j \mu^k = - \mu^k \mu^j$, $j, k, l = 1 \text{-} 3$.
To allow for the possibility that A may transform as a four-vector, we ask that it have four
components. A convenient hermitian representation for the $\mu^\alpha$ is then
\cite{Moses1,Moses2,Moses3}
\begin{align}\label{mualpha}
{\mu}^{0} &= \left ( {\begin{array}{cccc} 1&0&0&0  \\ 0&1&0&0
\\  0&0&1&0  \\ 0&0&0&1
\end{array}} \right ) = {\bf 1}, ~~\mu^{1}
 =  \left (\begin{array}{cccc} 0&-1&0 &0
\\  -1&0&0&0  \\  0&0&0&-i  \\  0&0&i&0 \end{array} \right ) ,  \nonumber
\\ \\
\mu^{2}  &=  \left (\begin{array}{cccc} 0&0&-1&0
\\  0&0&0&i  \\  -1&0&0&0  \\ 0&-i&0&0 \end{array} \right ) ,
~~\mu^{3}  =  \left (\begin{array}{cccc} 0&0&0&-1
\\  0&0&-i&0   \\  0&i&0&0
\\  -1&0&0&0 \end{array} \right ) \nonumber.
\end{align}
The $\mu^j$ of (\ref{mualpha}) relate to the vector rotation and boost
generators\footnote{\label{Ryder}${({\cal J}^j)^k}_l = -i\epsilon_{jkl}$, all other ${({\cal
J}^j)^\alpha}_\beta = 0$. ${({\cal K}^j)^0}_j = -i = {({\cal K}^j)^j}_0$, all other ${({\cal
K}^j)^\alpha}_\beta = 0$, see L. H. Ryder, {\it Quantum Field Theory\/} (Cambridge University
Press, Cambridge, 1996), p. 37; Ref. \onlinecite{Jax1}.} ${\cal J}^j$ and ${\cal K}^j$ via $\mu^j
= -\mu_j = {\cal J}^j-i{\cal K}^j$ and satisfy $\text{det}{\,}\mu^{j}=1$, $\text{tr}{\,}\mu^{j} =
0$, $\mu^{j}\mu^{k} = {i}\varepsilon_{jkl}\mu^{l} + \delta_{jk}{{\bf 1}}$, $j, k, l = 1 \text{-}
3$. They may be reduced to a block-diagonal composition of the Pauli
matrices\footnote{{\scriptsize{$\sigma^0 = \left (\begin{array}{cc}1&0 \\ 0&1\end{array} \right
)$, $~\sigma^{1} = \left (\begin{array}{cc} 0&1
\\ 1&0 \end{array} \right )$, $~\sigma^{2} = \left (\begin{array}{cc}0&-i \\
i&0\end{array} {\!}\right )$, $~\sigma^{3} = \left (\begin{array}{cc} 1&0 \\ 0&-1
\end{array} \right )$.}}} $\sigma^j$ by applying
the unitary transformation ($U^{-1} = U^{\dag}$)
\begin{equation}\label{U}
U {\!}\equiv{\!} {\frac{1}{2}}  \left (\begin{array}{cccc} 1&-1&i&-1
\\ 1&-1&-i&1
\\ i&i&1&-i \\ -i&-i&1&-i \end{array} \right ){\!}, ~~~~ \Sigma^j {\!}\equiv{\!} U \mu^j U^{\dag}
{\!}={\!} \left (\begin{array}{cc} \sigma^j & 0 \\ 0 & \sigma^j
\end{array} \right ){\!},
\end{equation}
$j=1,2,3$. We will refer to Eqs. (\ref{mualpha}) and (\ref{U}) as the R{\"u}mer matrices in the
{\it Minkowski\/} and {\it Pauli\/} bases respectively. In the Minkowski basis, we identify
\begin{gather}
{\rm A} = \left (\begin{array}{c} {\Phi}  \\  {A}_{x}
\\  {A}_y  \\  {A}_z \end{array} \right ), ~~~~~{\rm C}
= \left (\begin{array}{c} \Gamma  \\  C _{x}  \\  C_y  \\  C_z
\end{array} \right )
= \left (\begin{array}{c} \Gamma \\  -E _{x} - i B _{x}
\\  -E_y - i B_y  \\  -E_z - i B_z
\end{array} \right ), \nonumber
\\ \label{ACY} \\
{\rm Y} = \left (\begin{array}{c} \Upsilon
\\  Y _{x}  \\  Y_y  \\  Y_z \end{array} \right ) = \left (\begin{array}{c} \rho +
\partial _{0} \Gamma  \\  J_{x}  -
\partial _{x} \Gamma \\  J_y  - \partial
_{y} \Gamma  \\  J_z  - \partial _{z} \Gamma
\end{array} \right ). \nonumber
\end{gather}
For A real, employing (\ref{mualpha}) to write out (\ref{MME}a) and (\ref{MME}b) yields the
gauge-fixing and field-potential (Gauss-Kirchhoff) relations (\ref{MME}a) \cite{GaussK} and
Maxwell's equations (\ref{MME}b) respectively\footnote{Bold Roman fonts indicate the ``spatial''
(1-3) portions of a given quantity, e.g., ${\bf A} = ({A}_x ,{A}_y ,{A}_z )$, however, these {\it
need not transform\/} as either a three-vector or as the spatial portions of a four-vector, as we
will see below.}
\begin{subequations}\label{PFME}
\begin{gather}
\left . \begin{array}{cc}  \partial_{0}{\Phi}+ {\bm{\nabla \cdot}{\bf A}} = \Gamma , \\
{\bm{\nabla \times}{\bf A}} = {\bf B}, \qquad -{\bm{\nabla}}{\Phi}-{\partial}_{0}{\bf A} = {\bf
E}, \end{array} \right \} \\ \left .
\begin{array}{ll} \bm{\nabla \cdot} {\bf E} = \rho, & \qquad {\bm{\nabla \times}{\bf B}} -
{\partial}_{0} {\bf E} = {\bf J}, \\ {\bm{\nabla \cdot}{\bf B}} = 0, & \qquad \bm{\nabla
\times}{\bf E} + {\partial}_{\rm 0} {\bf B} = 0. \end{array} \right \}
\end{gather}
\end{subequations}
For A complex, the situation is slightly more involved. We may independently define ${\bf B}
\equiv {\bm{\nabla \times}{\bf A}}$ and ${\bf E} \equiv -{\bm{\nabla}}{\Phi}-{\partial}_{0}{\bf
A}$, under which Eqs. (\ref{PFME}) remain valid as they are. {\bf E} and {\bf B} are then also
complex and no longer the real and imaginary parts of {\bf C} \cite{Cref}. However, for complex A,
let us define alternate electric and magnetic fields as the real and imaginary parts of {\bf C}
and equate real and imaginary parts of Eqs. (\ref{MME}). In shorthand notation, let
\begin{gather}
\left ( \begin{array}{c} \Phi \\ {\bf A}
\end{array} \right ) = \left ( \begin{array}{c} \phi+i\theta \\ {\bf a}+i{\bf b} \end{array}
\right ), ~~~ \left ( \begin{array}{c} \Gamma \\ {\bf C}
\end{array} \right ) = \left ( \begin{array}{c} \gamma+i\delta \\ -{\bm{\mathcal E}}
-i{\bm{\mathcal B}} \end{array} \right ), \nonumber \\  \rule{0mm}{7mm}\left (
\begin{array}{c} \rho \\ {\bf J}
\end{array} \right ) = \left ( \begin{array}{c} \epsilon+i\beta \\ {\bf j}+i{\bf m} \end{array}
\right ),  \label{ImA}
\end{gather}
where $\phi$, $\theta$, ${\bf a}$, ${\bf b}$, $\gamma$, $\delta$, ${\bm{\mathcal E}}$,
${\bm{\mathcal B}}$, $\varepsilon$, $\beta$, ${\bf j}$, ${\bf m}$ are real. Equations
(\ref{MME}a) and (\ref{MME}b) respectively then become
\begin{subequations}\label{PFME2}
\begin{gather}
\left . \begin{array}{c} \partial_{0}{\phi}+ \bm{\nabla \cdot}{\bf a} = \gamma , \qquad
\partial_{0}{\theta}+ \bm{\nabla \cdot}{\bf b} = \delta , \\
\begin{array}{r}
\bm{\nabla \times}{\bf a} - \bm{\nabla}\theta - \partial_0 {\bf b} = \bm{\mathcal B}, \\ -
\bm{\nabla \times}{\bf b} - \bm{\nabla}\phi -\partial_0 {\bf a} = \bm{\mathcal E},
\end{array}
\end{array} \right \} \\ \left .
\begin{array}{ll} \bm{\nabla \cdot} {\bm{\mathcal E}} = \varepsilon, &
\qquad \bm{\nabla \times}\bm{\mathcal B} - \partial_0 \bm{\mathcal E} = {\bf j},
\\ \bm{\nabla \cdot}\bm{\mathcal B} = \beta, & \qquad \bm{\nabla \times}\bm{\mathcal E}
+ \partial_0 \bm{\mathcal B} = -{\bf m}.
\end{array} \right \}
\end{gather}
\end{subequations}
This is the {\it standard magnetic charge-density current\/} form of Maxwell's equations
\cite{Dirac,Jax2,magsour}. As usual, if the ratio of electric to magnetic charge is the same for
all sources, this form may be returned to the electric-source-only Maxwell equations (\ref{PFME})
by a duality transformation \cite{Jax2}. Duality transformations here are simply phase rotations,
$e^{i\lambda}$ applied to A, C, and Y. Thus, if a phase angle $\lambda$ exists such that $\rho$
and ${\bf J}$, or equivalently $\Phi$ and {\bf A}, can be made real, then Eqs. (\ref{PFME2}) are
the traditional Maxwell equations (\ref{PFME}) under phase rotation.

Operating on (\ref{MME}b) with $\mu^\alpha \partial^\alpha$ yields the charge conservation and
electromagnetic wave equations
\begin{subequations}\label{cont}
\begin{align}
&\partial_0 \rho + {\bm{\nabla \cdot}{\bf J}} = 0, \\ &\partial^\alpha \partial_\alpha ({\bf
E}+i{\bf B}) = {\bm{\nabla}}\rho +\partial_0 {\bf J} - i{\bm{\nabla \times}{\bf J}}
\end{align}
\end{subequations}
for $\rho$, {\bf J}, {\bf E}, and {\bf B} real or complex.

From here forward, we assume Eqs. (\ref{MME}) are in four-dimensional form and take these to be
our fundamental expression of the Maxwell, Gauss-Kirchhoff, and gauge-fixing equations
\cite{MME*}.

\section{\label{sec:trans}SPIN-1/2 AND SPIN-1 TRANSFORMATIONS OF THE MAXWELL VARIABLES}

A and Y may transform together under any four-dimensional representation of the Lorentz group and
leave Eq. (\ref{AY}) alone covariant. But restricting our attention to (\ref{AY}) comes at the
expense of identifying and transforming the field equations proper (\ref{MME}b). Similarly,
setting ${\rm Y} = 0$ (gradient-constrained sources) allows C to transform under any
four-dimensional representation of the Lorentz group and leave Eq. (\ref{MME}b) alone covariant.
The constrained-source equations of Simulik and Krivsky \cite{Sim1,Sim2,Sim3}, and the massless
equations of Darwin \cite{Darwin}, Madelung \cite{Made}, Ljolje \cite{Lj}, and Bruce \cite{Bruce},
correspond to writing (\ref{MME}b) by itself under this ${\rm Y} = 0$ condition (in a Dirac matrix
form) and treating C as a Dirac spinor. Requiring $\Gamma = 0$ in all frames returns Moses's
Maxwell equations \cite{Moses1,Moses2,Moses3}, forces the Lorentz gauge, and equates Y with the
current $(\rho, {\bf J})$, guaranteeing that the Maxwell variables will transform under their
usual spin-1 rules \cite{Moses1}. With both ${\rm Y} = 0$ and $\Gamma = 0$ in all frames (i.e.,
the Lorentz gauge and no sources), Eq. (\ref{MME}b) alone reduces to Maxwell's equations in a
photon wave-equation form, again spin-1, often pared to a $3\times3$ equation only (see
\onlinecite{guide} and \onlinecite{Cref}). But to find their general Lorentz transformation
properties, we will impose covariance on Eqs. (\ref{MME}) {\it without\/} restrictions on A, C and
Y.

Transforming coordinates as usual between frames {\it K\/} and {\it K\/}$'$ via $x'^{\beta} =
{\ell^{\beta}}_{\alpha}x^{\alpha}$ \cite{Brown}, with $g_{{\alpha}{\beta}}dx^{\alpha}dx^{\beta} =
dx_{\beta}dx^{\beta}$ and thus ${\ell^{\beta}}_{\sigma}{\ell_{\beta}}^{\alpha} =
{\delta^{\alpha}}_{\sigma}$, we require in {\it K\/}$'$
\begin{subequations}\label{MME'}
\begin{align} \mu'^{\alpha}
{\partial'} ^{\alpha} {\rm A'}({x}') &= {\rm C'}({x}') \label{MME'a}
\\ \mu'^{\beta} {\partial'} _{ \beta } {\rm C'}({x}') &=
{\rm {Y'}}({x}'), \label{MME'b}
\end{align}
\end{subequations}
where we take $\mu'^\alpha = \mu^\alpha$. From (\ref{AY}), A and Y must transform under the
same rule, so we seek transformation matrices ${\cal R}$ and ${\cal S}$ satisfying
\begin{gather}
{\rm A'} (x') = {\cal R}(\ell){\rm A}(x), ~~ {\rm Y'} (x') = {\cal R}(\ell){\rm Y}(x),
\nonumber\\{\rm C'} (x') = {\cal S}(\ell){\rm C}(x).\label{RST}
\end{gather}
As we show in Appendix \ref{sec:covar}, Eqs. (\ref{MME'}) and (\ref{RST}) have two solutions for
${\cal R}$ and ${\cal S}$. The first is
\begin{subequations}\label{RS1/2}
\begin{align}
{\cal R}(\bm{\varphi}, \bm{\zeta}) &= \exp \left [ i(\bm{\varphi} + i \bm{\zeta}) \bm{\cdot}
{\frac{\bm{\mu}}{2}} \right ] \equiv R(\bm{\varphi}, \bm{\zeta})
\\ {\cal S}(\bm{\varphi}, \bm{\zeta}) &= \exp \left [
i(\bm{\varphi} - i \bm{\zeta}) \bm{\cdot} {\frac{\bm{\mu}}{2}} \right ] \equiv S(\bm{\varphi},
\bm{\zeta}),
\end{align}
\end{subequations}
and the second
\begin{subequations}\label{RS1}
\begin{align}
{\cal R}(\bm{\varphi}, \bm{\zeta}) &= \exp [ i(\bm{\varphi} \bm{\cdot} \bm{{\cal J}} + \bm{\zeta}
\bm{\cdot} \bm{{\cal K}}) ] \\&= R(\bm{\varphi}, \bm{\zeta})R^{\textstyle{\ast}}(\bm{\varphi},
\bm{\zeta}) \equiv L(\bm{\varphi}, \bm{\zeta}),\nonumber
\\ {\cal S}(\bm{\varphi}, \bm{\zeta}) &= \exp [ i(\bm{\varphi} - i \bm{\zeta}) \bm{\cdot}
\bm{{\cal J}} ] \\&= S(\bm{\varphi}, \bm{\zeta})R^{\textstyle{\ast}}(\bm{\varphi}, \bm{\zeta})
\equiv M(\bm{\varphi}, \bm{\zeta})\nonumber
\end{align}
\end{subequations}
(* = complex conjugate), where $\bm{\varphi}$ and $\bm{\zeta}$ are the usual rotation and boost
parameters respectively. Note that covariance of Maxwell's equations (\ref{MME}b) alone, or
similarly (\ref{MME}a) alone, also leads to solutions (\ref{RS1/2}) and (\ref{RS1}) for ${\cal
R}$ and ${\cal S}$. We provide the {\it improper\/} transformations of A, C, and Y at the end
of Appendix \ref{sec:covar}.

$\bm{{\cal J}}$ and $\bm{{\cal K}}$ of (\ref{RS1}) are the vector rotation and boost generators
composing matrices (\ref{mualpha}), and thus $L$ corresponds to the $(1/2,1/2)$ representation of
the Lorentz group ($L = [{\ell^\alpha}_\beta]$) and $M$ the scalar-tensor $(0,0)\oplus(0,1)$
representation \cite{Brown}. Equations (\ref{RS1}) together thus return the expected spin-1
transformation of the Maxwell variables.

To understand the transformation (\ref{RS1/2}), consider the $\mu^j$ in the Pauli basis, as the
$\Sigma^j$ of Eq. (\ref{U}). We will mark objects in this basis with a dot underneath. Applying
the unitary transformation (\ref{U}), $R$ and $S$ become
\begin{align}\label{dotRS1/2}
\text{\d{\it R\/}}(\bm{\varphi}, \bm{\zeta}) &= \exp \left [ i(\bm{\varphi} + i \bm{\zeta})
\bm{\cdot} {\frac{\bm{\Sigma}}{2}} \right ] {\,}(\text{under-dot} {\!}={\!} \text{Pauli basis})
\nonumber
\\ \text{\d{\it S\/}}(\bm{\varphi}, \bm{\zeta}) &= \exp \left [
i(\bm{\varphi} - i \bm{\zeta}) \bm{\cdot} {\frac{\bm{\Sigma}}{2}} \right ] = \text{\d{\it
R\/}}^{-1\dag}(\bm{\varphi}, \bm{\zeta}),
\end{align}
and thus the top two and bottom two components of $\text{\d{\rm A}} = U {\rm A}$ and $\text{\d{\rm
Y}} = U {\rm Y}$ transform separately as $(1/2,0)$ two-spinors, while those of $\text{\d{\rm C}} =
U {\rm C}$ transform as $(0,1/2)$ two-spinors \cite{Brown}. $R$ and \d{\it R\/} are thus
equivalent members of the $(1/2,0)\oplus(1/2,0)$ representation of the Lorentz group, while $S$
and \d{\it S\/} belong to their conjugate $(0,1/2)\oplus(0,1/2)$ representation. $R$ and $S$ thus
perform a spin-1/2 transformation of A, C, and Y and consequently of the Maxwell field and
current. (See Appendix \ref{sec:covar}.)

Since $R$ and $S$ are complex, even if the spin-1/2 A, {\bf E}, {\bf B}, $\rho$, and {\bf J}
are real in one frame, or a duality transformation will eliminate their imaginary parts in that
frame, they will remain complex in other frames. Interpreting spin-1/2 field and source terms
as real quantities in all frames therefore requires the Dirac magnetic current form of
Maxwell's equations (\ref{PFME2}).

\section{\label{sec:four}THE MINKOWSKI SPINOR}

\begin{table*} \caption{\label{Table}Index notation for vector-like representations
of the Lorentz group, $R^T g R = S^T g S = L^{T} g L = M^{T} g M = g$.}
\begin{ruledtabular}
\begin{tabular}{llll}
Lorentz Group Representation & Group Element & Alphabetic Index & Numeric Index
\\ \hline $(1/2,0)\oplus(1/2,0)$ & $R$ (\text{\d{\it R\/}}) \rule{0mm}{3.5mm}
& {\it a}, {\it b}, {\it c}, ..., & {\it 0}, {\it 1}, {\it 2}, {\it 3}
\\ $(0,1/2)\oplus(0,1/2)$ & $S$ (\text{\d{\it S\/}}),
$R^{\textstyle{\ast}}$ ($\text{\d{\it R\/}}^{\textstyle{\ast}}$) & ${\it {\dot a}}$, $\it{\dot
b}$, $\it{\dot c}$, ..., & $\it{\dot 0}$, $\it{\dot 1}$, $\it{\dot 2}$, $\it{\dot 3}$ \\
$(1/2,1/2)$ & $L$ (\text{\d{\it L\/}}) \rule{0mm}{3.5mm} & $\alpha$, $\beta$, $\gamma$, ..., & 0,
1, 2, 3 \\ $(0,0)\oplus(0,1)$ & $M$ (\text{\d{\it M\/}}) & $\hat{\alpha}$, $\hat{\beta}$,
$\hat{\gamma}$, ..., & $\hat{0}$, $\hat{1}$, $\hat{2}$, $\hat{3}$
\\ Undefined transformation properties & --- &${\widetilde \alpha}$, ${\widetilde \beta}$,
${\widetilde \gamma}$, ..., \rule{0mm}{3.5mm} &${\widetilde 0}$, ${\widetilde 1}$, ${\widetilde
2}$, ${\widetilde 3}$
\end{tabular}
\end{ruledtabular}
\end{table*}

Before examining the spin-1/2 Maxwell invariants, we require a brief investigation of the
properties of {\it R\/} and {\it S\/}. For representations of the Lorentz group over a complex
domain, two types of invariant norms may result \cite{Barut,Craw}. The first involves a vector and
its complex conjugate and the second the simple inner product without complex conjugation. The
first implies $R^{\dag} \kappa R = \kappa$, where $\kappa$ is the metric. But as in the $(1/2,0)$
representation, $\kappa$ has no nonzero solution for $R \in (1/2,0)\oplus(1/2,0)$ [nor for $S \in
(0,1/2)\oplus(0,1/2)$].\footnote{From Appendix \ref{sec:covar}, considering small rotations and
boosts separately, ${\cal R}_{\rm R}^{\dag} \kappa {\cal R}_{\rm R} = \kappa$ implies
$\kappa\mu^j=\mu^j \kappa$, while ${\cal R}_{\rm B}^{\dag} \kappa {\cal R}_{\rm B} = \kappa$
implies $\kappa\mu^j=-\mu^j \kappa$, hence $\kappa=0$.} We thus look for a norm-squared of the
second type and therefore a solution to $R^T \kappa R=\kappa$.

In the Pauli basis, such complex norms-squared may be formed by dotting each top and bottom
two-spinor with itself, or each with the other, using the two-spinor metric $\sigma^2$.
However, taking our spinor components to be classical commuting functions, as opposed to
Grassmann variables, only dotting top and bottom two-spinors with each other will return a
nonzero squared-magnitude. Let $\text{\d{Z}} = \text{col}(\text{\d{$\chi$}},\text{\d{$\xi$}})$
represent a classical 4-spinor transforming via $\text{\d{\it R\/}}$, with $\text{\d{$\chi$}}$
and $\text{\d{$\xi$}}$ its top and bottom two-spinor segments respectively. In general, a
nonzero complex norm-squared results from the metric
\begin{equation}\label{dg}
\text{\d{\it g\/}} \equiv \left ( \begin{array}{cc} 0 & -\sigma^2 \\ \sigma^2 & 0
\end{array} \right )
\end{equation}
and inner product
\begin{equation}\label{ZgZ}
\text{\d{Z}}^T \text{\d{\it g\/}} \text{\d{Z}} = \text{\d{$\xi$}}^T \sigma^2 \text{\d{$\chi$}}
- \text{\d{$\chi$}}^T \sigma^2 \text{\d{$\xi$}} = 2\text{\d{$\xi$}}^T \sigma^2
\text{\d{$\chi$}}
\end{equation}
(similarly for \d{\it S\/}). In fact, $(1/2,0)\oplus(1/2,0)$ and $(0,1/2)\oplus(0,1/2)$ are the
only spin-1/2 representations of the Lorentz group for which 4-spinor squared-magnitudes of the
type $\psi^T \kappa \psi$, for $\psi$ classical, may be nonzero. This is because the only other
4-spinor representations are $(1/2,0)\oplus(0,1/2)$ and its conjugate, both of which form such
contractions by dotting each top and bottom two-spinor with itself, e.g. $\psi^T \Sigma^2
\psi$, which is zero for classical $\psi$. Instead, the only nonzero classical Dirac spinor
norm-squared is of the first type, $\psi^\dag \gamma^0 \psi$.\footnote{\label{gamma}Equations
(\ref{2DE}) below.} However, with anticommuting Grassmann components, $\text{\d{Z}}^T
\text{\d{\it g\/}} \text{\d{Z}}=0$ while $\text{\d{Z}}^T \Sigma^2 \text{\d{Z}}$, $\psi^T
\Sigma^2 \psi$, and $\psi^\dag \gamma^0 \psi$ are nonzero. {\it R\/} and {\it S\/} with the
metric \d{\it g\/} thus describe 4-spinors with commuting variables and simple norms-squared.
They are the lowest dimensional representations of the Lorentz group with nonzero spinor norms
of this type.

The metric (\ref{dg}) has a further remarkable property. In the Minkowski basis, \d{\it g\/} is
the {\it Minkowski metric\/}
\begin{equation}\label{g}
g= U^T \text{\d{\it g\/}} U = \left ( \begin{array}{cccc} 1&0&0&0 \\ 0&-1&0&0 \\ 0&0&-1&0 \\
0&0&0&-1
\end{array} \right ).
\end{equation}
The invariant (\ref{ZgZ}) thus takes the {\it vector-like\/} form $\text{\d{Z}}^T \text{\d{\it
g\/}} \text{\d{Z}}{\,}$\linebreak[0]$={\rm Z}^T g {\rm Z}$. We can obtain the same result from the
Minkowski basis directly. Letting $i(\bm{\varphi} + i \bm{\zeta}) \bm{\cdot} \bm{\mu}/2 \equiv
\pi$, we have \cite{Jax1}
\begin{equation}
gR^{T}g = e^{g\pi^{T}g},
\end{equation}
and employing the relation $g {\mu^{j\/ T}} g = -\mu^{j}$, $j = 1, 2, 3$, we find $g\pi^{T}g =
-\pi$ and thus
\begin{equation}\label{RgR}
R^{T} g R = g.
\end{equation}
Similarly $S^{T} g S = g = M^{T} g M = L^{T} g L$, or equivalently
\begin{equation}\label{dRgRs}
\text{\d{\it R\/}}^T \text{\d{\it g\/}} \text{\d{\it R\/}} = \text{\d{\it S\/}}^T \text{\d{\it
g\/}} \text{\d{\it S\/}} = \text{\d{\it M\/}}^{T} \text{\d{\it g\/}} \text{\d{\it M\/}} =
\text{\d{\it L\/}}^{T} \text{\d{\it g\/}} \text{\d{\it L\/}} = \text{\d{\it g\/}}{\,}.
\end{equation}
{\it Invariant inner products formed in the\/} $(1/2,0)\oplus(1/2,0)$ {\it or\/}
$(0,1/2)\oplus(0,1/2)$ {\it representations of the Lorentz group\/}, like those of $(1/2,1/2)$,
$(0,0)\oplus(0,1)$ and its conjugate \cite{0001}, {\it thus take the Minkowski form\/}
\begin{equation}\label{XgZ} {\rm X}^T g {\rm Z} = \text{\d{X}}^T \text{\d{\it g\/}}
\text{\d{Z}}~,
\end{equation}
where X and Z, or $\text{\d{X}}={\it U}{\rm X}$ and $\text{\d{Z}}={\it U}{\rm Z}$, transform
together under any one of these representations.

Accordingly, we introduce new notation for {\it vector-like\/} representations of the Lorentz
group, {\it those which form invariants with the Minkowski metric.\/} Let X be a column vector
transforming via any Lorentz group representation ${\sf V}$ satisfying ${\sf V}^Tg{\sf V}=g$. Thus
${\sf V}$ may belong to any one of the representations $(1/2,0)\oplus(1/2,0)$,
$(0,1/2)\oplus(0,1/2)$, $(1/2,1/2)$, $(0,0)\oplus(0,1)$, or $(0,0)\oplus(1,0)$, designated here by
$R$, $S$ ($R^{\textstyle{\ast}}$), $L$, $M$, $M^{\textstyle{\ast}}$ respectively \cite{reps}. Now
define the Riemannian conjugate
\begin{equation}\label{Xbar}
\overline{\rm X} \equiv (g{\rm X})^T = {\rm X}^T g.
\end{equation}
$\overline{\rm X}$ is thus a four-component row of the covariant components of X. This notation
does not distinguish between vector-like Lorentz group representations ${\sf V}$. To
distinguish between these representations we will employ index notation using alphabetic and
numeric indices with different fonts, each indicating a separate representation of the Lorentz
group. Our index notation is summarized in Table \ref{Table} and further described in Appendix
\ref{sec:nota}.

A four-component spinor ${\rm Z}$ transforming via $R$, with components $Z^a$, $a = {\it
0}\text{-}{\it 3}$, and its covariant counterpart $g{\rm Z}$, transforming via $(R^{T})^{-1}$,
with components $Z_a = g_{a b}{Z}^{b}$ $(g_{ab} = g_{\alpha \beta})$, satisfy
\begin{gather}
{Z'}^a = {R^a}_b{\,} Z^{b}, ~~~~~~~~ {Z'}_a = {R_a}^b{\,} Z_b {\,}, \nonumber
\\ \overline{{\rm Z}}' {\rm Z}' = {Z'}^a {Z'}_a = {R^a}_b {R_a}^c Z^b Z_c = Z^b Z_b{\,} =
\overline{{\rm Z}} {\rm Z},
\end{gather}
where ${R^a}_{b} {R_a}^{c} = {\delta_{b}}^{c}$ ($= 1$ for $b = c$, 0 for $b \neq c$). ${{\rm
Z}}^{\textstyle \ast}$, with components ${Z}^{a {\textstyle \ast}} \equiv {Z}^{\dot{a}}$ ({\it
a\/}-dot), transforms via $R^{\textstyle {\ast}}$. Its covariant counterpart ${g\/}{{\rm
Z}}^{{\textstyle \ast}}$, with components $Z_{\dot{a}} = g_{\dot{a} \dot{b}} Z^{\dot{b}}$
[$g_{\dot{a} \dot{b}} \equiv g_{ab}^{{\textstyle \ast}} = g_{ab} $], thus transforms via $S =
(R^{\dag})^{-1}$ [see (\ref{SRdag})]. That is
\begin{gather}
g{{\rm Z}'}^{\textstyle \ast} = ({R ^{\dag } } ) ^{-1} (g {\rm Z} ^{\textstyle \ast} ) = S ( g
{\rm Z} ^{\textstyle \ast}), \nonumber \\ \text{with}~~~~~{{Z'} } ^{\dot{a} } = {R ^{\dot{a}}}
_{\dot{b}}{\,} Z^{\dot{b}} ~~~~\text{and}~~~~ {{Z'}} _{\dot{a}} = {R _{\dot{a}}} ^{\dot{b}}{\,}
Z_{\dot{b}} \label{xi*}
\end{gather}
[${R^{\dot{a}}}_{\dot{b}} = ({R^{a}}_{b})^{\textstyle \ast}$], and thus
\begin{equation}\label{xi*g}
\overline{{\rm Z}}'^{\textstyle \ast} {{\rm Z}}'^{\textstyle \ast} = {Z'} ^{\dot{a}} {Z'}
_{\dot{a}} = {R ^{\dot{a}}} _{\dot{b}} {R _{\dot{a}}} ^{\dot{c}} Z^{\dot{b}} Z_{\dot{c}} =
Z^{\dot{b}} Z_{\dot{b}} =\overline{{\rm Z}}^{\textstyle \ast} {{\rm Z}}^{\textstyle \ast},
\end{equation}
where ${R ^{\dot{a}}} _{\dot{b}} {R _{\dot{a}}} ^{\dot{c}} = {\delta _{\dot{b}}} ^{\dot{c}}$ ($=
{\delta _{b}}^{c}$) \cite{hispin}. We will refer to four-component objects transforming under the
$(1/2,0)\oplus(1/2,0)$ and $(0,1/2)\oplus(0,1/2)$ representations of the Lorentz group as {\it
Minkowski spinors\/}, or as simply {\it four-spinors\/}, due to their use of the Minkowski metric
and direct analogy with four-vectors.

\section{\label{sec:invar}SPIN-1/2 MAXWELL INVARIANTS}

Under spin-1/2 transformation, A and Y have contravariant Minkowski spinor components ${A}^{a}
= (\Phi, {\bf A})$ and ${Y}^{a} = (\Upsilon, {\bf Y})$, while those of C are covariant
components of a $(0,1/2)\oplus(0,1/2)$ Minkowski spinor and thus designated $C _{\dot{a}} = (C
_{\it{\dot 0}}, C _{\it{\dot 1}}, C _{\it{\dot 2}}, C _{\it{\dot 3}}) = (\Gamma, C_x , C_y ,
C_z )$. We may now form the primary invariant inner products of A, C, and Y in the spin-1/2
case. They are
\begin{subequations}\label{AgA}
\begin{align}
A^a A_a &= \overline{\rm A}{\rm A} = {\Phi}^2 - {\bf A}^{2}, \\ {C}^{\dot{a}} C_{\dot{a}} &=
\overline{\rm C}{\rm C} = \Gamma ^{2} - {\bf E}^{2} + {\bf B}^{2} - 2i {\bf E} \bm{\cdot} {\bf
B}, \\ Y^a Y_a &= \overline{\rm Y}{\rm Y} = (\rho +
\partial_0 \Gamma)^2 - ({\bf J} - \bm{\nabla}\Gamma)^2, \\
A^a Y_a &= \overline{\rm A}{\rm Y} = \Phi(\rho + \partial_0 \Gamma) - {\bf A}{\bm{\cdot}}({\bf
J} - \bm{\nabla}\Gamma).
\end{align}
\end{subequations}
{\it Clearly, each of these inner products is also an invariant of the spin-1 Maxwell field\/}.
{\it That is,\/} $\overline{\rm A}{\rm A}${\it ,\/} $\overline{\rm C}{\rm C}${\it ,\/}
$\overline{\rm Y}{\rm Y}${\it , and\/} $\overline{\rm A}{\rm Y}$ {\it are each invariant under
both spin-1 and spin-1/2 transformations of the Maxwell field and current.\/}

If we include the complex conjugates of A, C, and Y, then any contravariant-covariant
combination, summed over like indices $a = {\it 0}{\text{-}}{\it 3}$ or $\dot{a} = \it{\dot
0}{\text{-}}\it{\dot 3}$, will also be invariant. For example
\begin{subequations}\label{AgC}
\begin{align}
A^{\dot{a}} C_{\dot{a}} &= {\rm A}^{\dag}{\rm C} = \Phi ^{\textstyle \ast} {\Gamma} - {\bf
A}^{\textstyle \ast} \bm{\cdot}({\bf E} + {i}{\bf B}),
\\ Y^{\dot{a}} C_{\dot{a}} &= {\rm Y}^{\dag}{\rm C} = \Upsilon ^{\textstyle \ast} \Gamma
- {\bf Y}^{\textstyle \ast} \bm{\cdot}({\bf E} + i{\bf B}), \\ C^a A_a &= {\rm C} ^{\dag} {\rm
A} = (A^{\dot a}{C}_{\dot a})^{\textstyle \ast}, \\ C^a Y_a &= {\rm C} ^{\dag} {\rm Y} =
(Y^{\dot a}{C}_{\dot a})^{\textstyle \ast}
\end{align}
\end{subequations}
are all spin-1/2 Maxwell invariants. However, these have no spin-1 counterparts stemming from
the fact that $S = R^{-1\dag}$ does not hold for $L$ and $M$ of Eqs. (\ref{RS1}).

Finally, we may construct four-vectors and quadri-vectors $[(0,0)\oplus(0,1)]$ out of Minkowski
spinors. For example, the spin-1/2 A, C, and Y combine to form contravariant four-vectors ${\rm
A}^\dag \mu_\alpha {\rm A}$, ${\rm C}^\dag \mu^\alpha {\rm C}$, ${\rm Y}^\dag \mu_\alpha {\rm
Y}$, ${\rm A}^\dag \mu_\alpha {\rm Y}$, and quadri-vectors ${\rm A}^\dag \mu_\alpha {\rm C}$
and ${\rm Y}^\dag \mu_\alpha {\rm C}$, with similar results for any like-transforming
four-spinors.

\section{\label{sec:lagrangian}DUAL SPIN-1/2--SPIN-1 MAXWELL AND PROCA LAGRANGIAN DENSITIES}

A Lagrangian density for Maxwell equations similar to our (\ref{MME}b) has eluded previous
attempts at its discovery \cite{Gerst,Jena,Gerst2}. Giannetto \cite{Gian} has written a simple
Lagrangian density equivalent to describing (\ref{MME}b) alone without sources and under the
Lorentz condition (${\rm Y}=\Gamma=0$). In a related six-component formalism, Drummond \cite{Drum}
has discussed a duality-invariant Lagrangian density for the source-free Maxwell field, as have
Abreu and Hott \cite{Abreu,Chow}. Sources included, Gersten \cite{Gerst,Gerst2} has described
``pseudo-Lagrangians'' for Maxwell wave equations similar to (\ref{MME}b), while Ljolje \cite{Lj}
provides an extensive Lagrangian description of the constrained-source Maxwell equations of
Darwin, derived from a dual spin-1--spin-1/2 (Dirac spinor) Lagrangian density. (See also
\onlinecite{Sachs1} and \onlinecite{Sim3}.) But in related four-component and similar
three-component matrix forms of Maxwell's equations proper, {\it with\/} sources, to quote Jena
{\it et al.\/} \cite{Jena}, ``In neither of these two approaches do the equations seem to follow
from an Euler-Lagrange variational principle.''

The problem is finding an appropriate Lagrangian density for equations like (\ref{MME}b) alone.
By itself, (\ref{MME}b) and its hermitian conjugate result from varying ${\rm C}^\dag$ and C in
the Lagrangian density
\begin{equation}
{\cal L}_{\rm C} = \gamma_{\rm C}[{\rm C}^\dag \mu_\alpha \partial^\alpha {\rm C} - 2{\rm
C}^\dag {\rm Y} - (\partial^\alpha {\rm C}^\dag \mu_\alpha){\rm C} + 2{\rm Y}^\dag {\rm C}],
\end{equation}
where $\gamma_{\rm C}$ is a constant with units of length.\footnote{${\cal L}_{\rm C}$ is
analogous to the Lagrangian density for the generalized spinor Maxwell equations of Sachs and
Schwebel, see M. Sachs, {\it General Relativity and Matter\/}, Ref. \onlinecite{Sachs1}, p. 115.
The factor $\gamma_{\rm C}$ does not affect the field equations themselves, but without it, ${\cal
L}_{\rm C}$ and its derived conservation laws will be dimensionally incorrect. In the Sachs
formalism, the analogous constant ($g_M$, also with units of length) appears explicitly in
equations coupling matter and electrodynamics. See related comments by Sachs.} A similar
Lagrangian can be written for (\ref{MME}a) alone. However, ${\cal L}_{\rm C}$ is invariant only in
the {\it spin-1/2\/} case [see Eqs. (\ref{AgC})]. Nor can ${\cal L}_{\rm C}$ be summed with its
counterpart for Eq. (\ref{MME}a) to return both Eqs. (\ref{MME}) under variation. Instead, a
Lagrangian density for either Eqs. (\ref{MME}b) or (\ref{MME}a), allowing spin-1 transformations
of the Maxwell field and current, requires a description of both these equations
simultaneously.\footnote{The Schwinger Lagrangian for the tensor Maxwell field also
returns both the Maxwell and Gauss-Kirchhoff equations (\ref{MME}) (without the gauge-fixing
equation). See J. Schwinger {\it et al.\/}, {\it Classical Electrodynamics\/} (Perseus, Reading,
Mass., 1998), Sec. 9.2; J. Schwinger, {\it Phys. Rev.\/} {\bf 91}, 713 (1953); also Ref.
\onlinecite{Barut}, p. 103.} For this purpose we also require the Riemannian conjugates of Eqs.
(\ref{MME}). These are
\begin{subequations} \label{MMEconj}
\begin{align}
\partial^{\alpha} \overline{\rm A} \mu_{\alpha} &= \overline{\rm C}
\\ \partial^{\beta} \overline{\rm C} \mu^{\beta} &= \overline{\rm Y}.
\end{align}
\end{subequations}
We now introduce the Lagrangian density
\begin{align}\label{L}
{\cal L} = \frac{1}{2}[&\overline{\rm C} \mu^\beta \partial^\beta {\rm A} - \overline{\rm A}
\mu_\alpha \partial^\alpha {\rm C} + 2\overline{\rm A} {\rm Y} - \overline{\rm C}{\rm C}
\\ &+ (\partial^\alpha \overline{\rm A} \mu_\alpha) {\rm C} - (\partial^\beta \overline{\rm
C} \mu^\beta) {\rm A} + 2\overline{\rm Y} {\rm A} - \overline{\rm C}{\rm C}], \nonumber
\end{align}
where the terms on the second line are the Riemannian conjugates of those on the first. Varying
${\cal L}$ with respect to the dynamic field variables A, C, $\overline{\rm A}$, and
$\overline{\rm C}$ yields Eqs. (\ref{MME}) and their counterparts (\ref{MMEconj}). From Eqs.
(\ref{MME}), (\ref{AgA}), and (\ref{MMEconj}), $\bf{\cal L}$ is a scalar function under {\it
both spin-1 and spin-1/2 Lorentz transformations of its field components\/}.

Adding a Proca mass term $-m^2 \overline{\rm A} {\rm A}$ to the Lagrangian density (\ref{L})
gives
\begin{equation}\label{LP} {\cal L} - m^2 \overline{\rm A} {\rm A} \equiv {\cal L}_P
\end{equation}
and leads to the Proca Maxwell equations
\begin{subequations}
\label{PME}
\begin{align}
\mu^{\alpha} \partial^{\alpha} {\rm A} &= {\rm C},
\\ \mu^{\beta} \partial_{\beta} {\rm C} &= {\rm Y} - m^2 {\rm A},
\end{align}
\end{subequations}
and their Riemannian conjugate. As in the Maxwell case, the Proca Lagrangian density ${\cal L}_P$
and its associated equations of motion remain covariant under {\it both\/} spin-1 and spin-1/2
transformations (\ref{RS1/2}) and (\ref{RS1}) of their field terms. From the four-divergence of
(\ref{PME}b), we find, as usual (see Ryder,$^3$ p. 67),
\begin{equation}\label{JA} {\partial}_{0}{\rho} + \bm{\nabla \cdot}{\bf J} =
m^2({\partial}_{0}{\Phi} + \bm{\nabla \cdot}{\bf A}),
\end{equation}
and thus for $m \neq 0$ charge is conserved only under the Lorentz condition
${\partial}_{0}{\Phi} + \bm{\nabla \cdot}{\bf A}=0$. However, unlike the vector field, where
the Lorentz condition $\partial_\beta A^\beta=0$ is invariant and the continuity equation would
thus hold for all inertial observers, $\partial_\beta A^b=0$ is {\it not\/} invariant and
consequently there are no circumstances where charge is conserved in all frames for the
spin-1/2 Proca field. (This is {\it not\/} true for the spin-1/2 Maxwell field, see Eq.
(\ref{Jtilde}) below.) We will discuss further consequences of the Lagrangian formalism in a
later analysis.

\section{\label{sec:halfEM}LOSS OF GAUGE INVARIANCE, CHARGE INVARIANCE, AND LORENTZ FOUR-FORCE
IN THE SPIN-1/2 CASE}

Clearly, within a single frame, {\bf E} and {\bf B} are invariant under a gauge transformation
regardless of the Lorentz transformation properties of the potentials $(\Phi, {\bf A})$. This
can easily be seen by letting $(\Phi, {\bf A}) \rightarrow (\Phi -
\partial^0 \Lambda, {\bf A} + \bm{\nabla} \Lambda) \equiv (\Phi_\Lambda, {\bf A}_\Lambda)$ in
Eqs. (\ref{MME}), after which {\bf E} and {\bf B} remain unaffected in the usual manner. But in
the spin-1/2 case, the loss of invariance of the four-divergence of the potentials
$\partial_\beta A^b$ rules out an invariant gauge-fixing condition, and as may thus be
surmised, problems arise when we apply a spin-1/2 Lorentz transformation to the gauge
transformed potentials $(\Phi_\Lambda, {\bf A}_\Lambda)$.

In frame $K$, let ${\rm A}_\Lambda \equiv {\rm A} - \partial \Lambda$, with $\partial \equiv
\text{col}(\partial_0, -\partial_x ,$\linebreak[0]$ -\partial_y , -\partial_z)$, and let $\cal R$
stand for any Lorentz group representation transforming ${\rm A}_\Lambda$ from {\it K\/} to $K'$
giving
\begin{equation}
{\rm A}'_\Lambda = {\cal R}{\rm A}_\Lambda = {\cal R}{\rm A} - {\cal R}(\partial \Lambda) =
{\rm A}' - {\cal R}(\partial \Lambda).
\end{equation}
The term ${\cal R}(\partial \Lambda)$ is a gauge transformation in $K'$ only if there exists a
function $\Lambda'$ such that
\begin{equation}\label{RLam}
\partial' \Lambda' = {\cal R}(\partial \Lambda).
\end{equation}
In the vector case, ${\cal R}=L$ and (\ref{RLam}) is solved for $\Lambda$ a scalar function. But
for ${\cal R}=R$, a solution for $\Lambda'$ exists only if $\Lambda$ is at most a linear function
of the coordinates.\footnote{With ${\!}RR^{\textstyle{\ast}} \partial \Lambda'(x') {\!}={\!} R
[\partial \Lambda (x)]$, we have ${\!}\partial \Lambda' [x'(x)] {\!}={\!}
{R^{\textstyle{\ast}}}^{-1}[\partial \Lambda (x)]$. Writing out the second equation and cross
differentiating the components of $\partial \Lambda'$, then relating the various
${R^{\textstyle{\ast}}}^{-1}_{\alpha\beta}$ using Eqs. (\ref{gen1/2+}), leads to the condition
$\partial_\alpha \partial_\beta \Lambda = 0$, $\alpha$, $\beta = 0 \text{-} 3$.} Yet even with
this restriction, $\Lambda$ is still a different function in every frame (not a scalar function).

However, if we seek changes in A that leave ${\bm{\mathcal E}}$ and ${\bm{\mathcal B}}$ unaltered,
then significant freedom remains in the spin-1/2 potentials. Defining ${\rm G} \equiv
\text{col}(\gamma, {\bf G})$ and requiring
\begin{equation}\label{G} \mu^\alpha\partial^\alpha
{\rm G} \equiv 0,
\end{equation}
the substitution
\begin{equation}\label{A+G} {\rm A} \rightarrow {\rm
A}_{\rm G} \equiv {\rm A} + {\rm G} = (\Phi + \gamma, {\bf A} +
{\bf G})
\end{equation}
leaves C and thus ${\bm{\mathcal E}}$ and ${\bm{\mathcal B}}$ unchanged in all frames for both
spinor and vector fields. Note that G satisfies the Lorentz condition $\partial_\beta G^b = 0$ in
all frames. This is a special {\it pseudo\/}-invariant, formed between G and the
partial-derivative operator only (see Appendix \ref{sec:nota}). But again, (\ref{A+G}) is not a
gauge transformation and thus does not (generally) leave {\bf E} and {\bf B} unchanged.

As could be expected, charge-invariance is also lost in the spin-1/2 case. The volume integral
of $\rho$ when Y and C transform as four-spinors is
\begin{equation}\label{q1/2} q_{1/2} \equiv \int
{\rho {\,} d ^{3} x} = \int {(Y^{\it 0} - \partial^0 C_{\it{\dot
0}})d^3 x},
\end{equation}
which is clearly not a scalar since $Y^{\it 0} - \partial^0 C_{\it \dot{0}}$ is not the top
component of a four-vector. Instead, $q_{1/2}$ has no well-defined transformation property.
However, the continuity equation (\ref{cont}a) remains in effect. That is,
\begin{equation}\label{Jtilde}
\partial_\alpha J^{\widetilde{\alpha}} = 0 ~~~~~\text{for}~~~~~ J^{\widetilde{\alpha}} = Y^a -\partial^\alpha
C_{\it{\dot 0}},
\end{equation}
and thus $q_{1/2}$ is still {\it conserved\/} in each frame, even though it is not invariant.

Covariance of the Lorentz four-force is also lost under spin-1/2 rules. The Lorentz four-force
would appear as 
\begin{equation}\label{LF} dp^\alpha/d\tau \rightarrow
q_{1/2}[\partial^\alpha(u_\beta A^b) - dA^a/{d\tau}],
\end{equation}
where $\tau$ is proper time and $u_\beta$ the four-velocity \cite{Gold}. The left side of
(\ref{LF}) transforms as a four-vector while the right side does not transform as any one type of
object. Indeed, any attempt to describe a covariant Lorentz force on an hypothetical spin-1/2
charge faces the formidable hurdle that $q_{1/2}$ has an undefined transformation property. This
problem, and others in this section, stem from the fact that $A^a$ and $J^{\widetilde{\alpha}}$ do
not transform with the same rule as the coordinates, destroying the usual symmetry between
spacetime and the Maxwell variables so important in the spin-1 case. We will discuss an
interesting resolution to these problems in our concluding remarks.

\section{\label{sec:MKGM}EQUIVALENCE OF THE KLEIN-GORDON MAXWELL AND TWO DIRAC EQUATIONS}

Finally, consider Maxwell's equations under the condition that A satisfies the Klein-Gordon
equation, as if to describe a massive vector boson \cite{Aitch1}. This condition can be introduced
by setting ${\rm Y} = -m^2{\rm A}/2$ in the Lagrangian density (\ref{L}) or setting ${\rm Y}=0$ in
the Proca Lagrangian (\ref{LP}), the resulting Lagrangian density is the same in either case.
Letting ${\rm C} \equiv -im {\rm F}$, with $m$ constant, the corresponding
equations of motion are \cite{Dow1,Dvoe2} 
\begin{subequations}\label{MKGM}
\begin{align}
\mu^{\alpha} \partial ^{\alpha} {\rm A} &= -i m {\rm F}
\\ \mu^{\beta} \partial _{\beta} {\rm F} &= -i m {\rm A},
\end{align}
\end{subequations}
where ${\partial} ^{\alpha} {\partial } _{\alpha} {\rm A} = -m^2 {\rm A}$. These equations
possess the same dual spin-1/2--spin-1 Lorentz transformation property as Eqs. (\ref{MME}) and
(\ref{PME}). Equations (\ref{MKGM}) are equivalent to a single eight-component or two
four-component Dirac equations of mass {\it m\/}. With
\begin{equation} \Omega (x)
\equiv \left ( \begin{array}{c} {\rm A} (x)
\\  {\rm F}(x) \end{array} \right ) ~~~ \text{and} ~~~
\lambda^{\alpha} \equiv \left (\begin{array}{cc} {\bf{0}} & {\mu}^{\alpha}  \\ {\mu}_{\alpha} &
{\bf{0}}
\end{array} \right ) \label{Psi}
\end{equation}
(${\bf{0}}$ is the $4\times4$ zero matrix), Eqs. (\ref{MKGM}) become
\begin{equation}\label{8DE}
\lambda^{\alpha} \partial _{\alpha} \Omega (x) = - i m \Omega (x),
\end{equation}
where $\lambda^{{\alpha}}\lambda^{{\beta}} + \lambda^{{\beta}}\lambda^{{\alpha}} =
2{g\/}^{{\alpha}{\beta}}$ \cite{8x8}. Equation (\ref{8DE}) may be
reduced via 
\begin{equation}
W \equiv V {\,} U_8 , ~~~ V \equiv \left ( \begin{array}{cccc} {\it{1}}
&{\it{0}}&{\it{0}}&{\it{0}}
\\ {\it{0}}&{\it{0}}& {\it{1}} &{\it{0}}
\\ {\it{0}}& {\it{1}} &{\it{0}}&{\it{0}} \\ {\it{0}}&{\it{0}}&{\it{0}}&{\it{1}}
\end{array} \right ), ~~~ U_8 \equiv \left ( \begin{array}{cc} U & {\bf{0}} \\
{\bf{0}} & U \end{array} \right )
\end{equation}
($W^{-1} = W^\dag$, {\it U\/} of Eq. (\ref{U}), ${\it{1}}$ and ${\it{0}}$ are the $2\times2$ unit
and zero matrices) \cite{redux}. Applying $W$ to (\ref{8DE}), with 
\begin{equation} W \Omega(x)\equiv \left (
\begin{array}{c} \psi (x)
\\ \eta (x) \end{array} \right ),
\end{equation}
and separating into block diagonal parts returns \cite{Lj}
\begin{subequations}\label{2DE}
\begin{align}
\gamma ^{\mu} \partial _{\mu} \psi (x) &= - i m \psi (x)
\\ \gamma ^{\mu} {\partial} _{\mu} \eta (x) &= - i m \eta (x),
\end{align}
\end{subequations}
where the $\gamma^\mu$ are the Dirac matrices.\footnote{{\scriptsize{$\gamma^0 ={\!} \left
({\!}\begin{array}{cc} {\it 0}&{\it 1} \\ {\it 1}&{\it 0}\end{array} {\!}\right ){\!}$,
$\gamma^i ={\!} \left ({\!}\begin{array}{cc} {\it 0} & {\sigma^i} \\ -{\sigma^i} & {\it 0}
\end{array} {\!}\right ){\!}$, $i=1,2,3$.}}} The Klein-Gordon Maxwell (KGM) equations
(\ref{MKGM}) and Dirac equations (\ref{2DE}) are thus equivalent and may each describe either
two real {\it vector\/} fields, corresponding to real and imaginary parts of A, or two {\it
Dirac\/} fields, represented by $\psi$ and $\eta$ of (\ref{2DE}), all of the same mass.

\section{\label{sec:concl}CONCLUSION}

Surprisingly, the Maxwell, Gauss-Kirchhoff, and gauge-fixing equations are indeed covariant under
{\it spin-1/2\/} Lorentz transformations of the Maxwell variables. Though we have begun its
analysis here, much work remains to complete a description of the spin-1/2 case. We might expect
that a well-transforming conserved current would be fundamental to this description. But while
$J^{\widetilde{\alpha}}=(\rho, {\bf J})$ is indeed conserved, it has no well-defined
transformation rule, transforming with the four-gradient of $\Gamma$ as part of the four-spinor
$Y^a$. Y is composed of the traditional Maxwell input variables, the current and gauge-fixing
function (the four-gradient of the latter). But under spin-1/2 rules, we are not at liberty simply
choose the Lorentz gauge and equate $Y^a$ with $(\rho, {\bf J})$ since gauge transformations of
the spin-1/2 Maxwell field are not Lorentz invariant. Moreover, charge in the spin-1/2 case also
has no well-defined transformation rule. We may still treat the spin-1/2 Maxwell field within each
frame as being generated by the conserved charge-density current in that frame, but as this is not
a covariant treatment the more appropriate interpretation would evidently be that the spin-1/2
Maxwell field has no traditional source in the Maxwell sense and that the input variables Y
effectively separate into sources and gauge-fixing functions only in the spin-1 case.

Despite the obvious differences between the spin-1 and spin-1/2 Maxwell fields, their most
striking features are their similarities. In their form presented here, they share exactly the
same Lagrangian density, exactly the same field equations, adding a Proca mass term leaves the
transformation properties of both fields unchanged, they share invariant inner products
$\overline{\rm A}{\rm A}$, $\overline{\rm C}{\rm C}$, $\overline{\rm Y}{\rm Y}$, $\overline{\rm
A}{\rm Y}$, and, as these invariants imply, they share the Minkowski metric. This last fact, a
consequence of the Minkowski metric's association with the $(1/2,0)\oplus(1/2,0)$ and
$(0,1/2)\oplus(0,1/2)$ representations of the Lorentz group [Eqs. (\ref{RgR}) and
(\ref{dRgRs})], is suggestive of a deeper connection between spacetime and the Maxwell
equations. Clearly, any coordinate transformation forming invariants with the Minkowski metric
leaves invariant the interval $(x^0)^2-(x^1)^2-(x^2)^2-(x^3)^2$. In an effort to find all
coordinate and field transformations that leave Maxwell's equations covariant, consider then a
transformation of coordinates under four-spinor rules --- a complex fermionic spacetime
(reminiscent of fermionic coordinates in supersymmetry). In such a space, the four-divergence
of the spin-1/2 potentials $\partial_b A^b$ would again be invariant, allowing gauge
transformations of the spin-1/2 Maxwell field; the charge-density current would independently
transform as a four-spinor $J^b$, just as it transforms as a four-vector $J^\beta$ in a
four-vector space; the spin-1/2 charge $q_{1/2}$ (\ref{q1/2}) would thus be invariant too, and
consequently the Lorentz four-force (\ref{LF}) would regain covariance. Though we have only
begun to explore the spin-1/2 Maxwell field in a four-vector space, the fact that four-spinor
coordinates may lead to a covariant spin-1/2 form of electrodynamics, analogous to vector
electrodynamics, suggests a broader investigation into a Minkowski-spinor form of spacetime.

Further, the fact that the Maxwell field is covariant under both spin-1 and spin-1/2
transformations suggests that other traditionally-spin-1 quantities may also have fermionic
properties, especially since spin-1/2 invariants may involve the Minkowski metric. We might
imagine the other gauge vector fields to be candidates for such quantities, however their built-in
necessity for gauge interactions appears to preclude their spin-1/2 transformation. But if not
these, perhaps generalizations of the dual-transforming Lagrangian density (\ref{L}) will uncover
such fields. The importance of such a search lies in the possibility of a wider fermion-boson
field equivalence, beyond the spinor-vector--Maxwell and Dirac-KGM versions discussed above. That
is, by describing both vector and spinor fields, the Maxwell equations (\ref{MME}) and Lagrangian
density (\ref{L}) describe a type of unified spinor-vector field. If further such
fermion-boson--equivalent fields exist, then supersymmetry may not be the only type of
fermion-boson unification consistent with Lorentz transformations. This type of unification need
not conflict with a supersymmetry-extended Standard Model, but it may offer an alternate approach
to connecting fermion and boson fields, perhaps one capable of working within the Standard Model
framework.\footnote{Recall the fermion mass problem (see K. S. Babu, hep-ph/\linebreak[0]9210250;
C. D. Froggett, hep-ph/9504323; also J. Bordes {\it et al.\/}, {\it Phys. Rev. D\/} {\bf 65},
093006 (2002) (hep-ph/0111369) and references therein). Fermion-boson unification within the
Standard Model would necessarily give reason for inclusion of the fermion fields.} A wider search
for fields, or compositions of fields, with multiple transformation properties would thus seem
warranted.

Finally, the original motivation for this study was not the Maxwell equations themselves, but an
exploration of the hypothesis that matter may be understood as an electromagnetic phenomenon.
Similar hypotheses have surfaced at various points in physics history with electromagnetic
descriptions of the electron mass \cite{Pet}, EM-like descriptions of the Dirac field
\cite{light}, and recently with analogies between the vacuum and superconducting media in quantum
field theory of gauge vector bosons \cite{super}. We note here that with the KGM--2-Dirac equation
equivalence, a substantial portion of the matter spectrum may have as its foundation either spin-1
or spin-1/2 EM-like waves. The natural inference, not unlike that drawn from superconductor models
of the vacuum, is that a unified description of matter and electromagnetic phenomena may be
pursued under the hypothesis that matter waves are essentially a form of light traveling in a
diverse electromagnetic-like medium. Further development of this hypothesis will require placing
the spin-1/2 Maxwell field in a quantum-mechanical context, a task that remains to be completed.


\begin{acknowledgments}
The author would like to express his deepest appreciation to Dr. Jose L. Balduz, Jr. of Mercer
University for numerous invaluable discussions on this subject leading to significant
improvements in the above presentation. The author would like to further thank the Mercer
University library for allowing him generous and long-term use of their materials; the
libraries of the Georgia Institute of Technology and the University of Georgia at Athens, and
the people and government of the state of Georgia, for providing public access to professional
literature in the field of physics through the University System of Georgia Libraries; Drs. R.
D. Peters, D. T. Young, M. J. Marone, A. Gsponer, P. Milonni, A. Gersten; the Physics
Departments of Mercer and Auburn Universities; the IBM Corporation; Drs. R. S. and M. A.
Armour, Sr.; and R. H. and K. K. Armour. The author extends a special acknowledgment to John
St. Clair Cline, IV.
\end{acknowledgments}

\appendix

\section{\label{sec:covar}COVARIANCE OF THE MAXWELL EQUATIONS}

Inserting the inverses of (\ref{RST}) into (\ref{MMEb}) and comparing with (\ref{MME'b}) we find
\cite{Brown}
\begin{equation}\label{Wmu}
{\cal R}(\ell) \mu^{\sigma} {\cal S}^{-1} (\ell) {\ell^{\beta}}_{\sigma} = \mu'^{\beta} =
\mu^{\beta}.
\end{equation}
Beginning with the infinitesimal case ${\ell^{\beta}}_{\alpha} = {\delta^{\beta}}_{\alpha} +
{\Delta\zeta^{\beta}}_{\alpha}$ we employ the expansions
\begin{equation}\label{RS}
{\cal R} = {\bf 1} + {\frac{i}{2}} \rho _{\alpha \beta } \Delta \zeta ^{\alpha \beta }~~~ {\rm
and} ~~~{\cal S} = {\bf 1} + {\frac{i}{2}} \sigma _{\alpha \beta } \Delta \zeta ^{\alpha
\beta},
\end{equation}
where the $\Delta\zeta^{jk} = -\epsilon_{jkl} \Delta \varphi^l$ parameterize rotations and the
$\Delta\zeta^{0j} = -\Delta\zeta^{j0} = - \Delta\zeta^j$ parameterize boosts, with
$\Delta\varphi^l$ the infinitesimal rotation angle and $\Delta\zeta^j$ the infinitesimal boost
angle, $j, k, l = 1\text{-}3$. From (\ref{Wmu}), we may also write
$\mu^{\alpha}{{\ell}^{\beta}}_{\alpha} = {\cal R}^{-1} ({\ell}) \mu^{\beta} {\cal S}({\ell})$,
and inserting ${\cal R}^{-1}$ and $\cal S$ gives
\begin{align}\label{aapprox}
\mu^{\alpha} {{\ell} ^{\beta}}_{\alpha} &= ({{\delta } ^{\beta} } _{\alpha} + \Delta {\zeta
^{\beta} } _{\alpha} ) \mu^{\alpha} = \mu^{\beta} + \Delta {\zeta ^{\beta} } _{\alpha}
\mu^{\alpha} \\ &= [{\bf 1} - i\rho _{ \gamma\nu } \Delta \zeta ^{ \gamma\nu } /2 ]\mu^{\beta}
[{\bf 1} + i\sigma _{ \gamma\nu } \Delta \zeta ^{ \gamma\nu } /2 ] \nonumber
\\ &\approx \mu^{\beta} + i\mu^{\beta} \sigma _{
\gamma\nu } \Delta \zeta ^{ \gamma\nu } /2 - i\rho _{ \gamma\nu } \Delta \zeta ^{\gamma\nu }
\mu^{\beta} /2 . \nonumber
\end{align}
Then, from $\Delta\zeta^{{\alpha}{\beta}}=-\Delta\zeta^{{\beta}{\alpha}}$ and manipulation of
indices,
\begin{equation}\label{delz}
\Delta {\zeta ^{\beta} } _{\alpha} \mu^{\alpha} = {\frac{1}{2}} \Delta \zeta ^{ \gamma\nu }
({{\delta } ^{\beta} } _{\gamma} \mu_{\nu } - {{\delta } ^{\beta} } _{\nu } \mu_{\gamma} ) .
\end{equation}
Combining (\ref{delz}) with (\ref{aapprox}) yields
\begin{equation}\label{musig}
\mu^{\beta} \sigma _{ \gamma\nu } - \rho _{ \gamma\nu } \mu^{\beta} =  i({{\delta } ^{\beta} }
_{\nu } \mu_{\gamma} - {{\delta } ^{\beta} } _{\gamma} \mu_{\nu } ) .
\end{equation}

Equations (\ref{musig}) have two sets of solutions for the generators $\rho_{{\gamma}{\nu}}$
and $\sigma_{{\gamma}{\nu}}$. The first is
\begin{align}\label{gen1/2}
\rho^{(1/2)}_{\gamma\nu} &= {\frac{i}{4}} [ \mu^{\gamma} , ~\mu^{\nu } ] - {\frac{i}{2}}({{\delta
} ^{0}}_{\gamma} \mu^{\nu} - {{\delta} ^{0}}_{\nu} \mu^{\gamma} ) , \nonumber \\
\sigma^{(1/2)}_{\gamma\nu} &= {\frac{i}{4}} [ \mu^{\gamma} , ~\mu^{\nu} ] + {\frac{i}{2}}
({{\delta} ^{0}}_{\gamma} \mu^{\nu} - {{\delta} ^{0}}_{\nu} \mu^{\gamma}),
\end{align}
or explicitly $\rho^{(1/2)}_{00} = \sigma^{(1/2)}_{00} = {{\bf 1}}$,
\begin{align}\label{gen1/2+}
\rho^{(1/2)}_{0j} = -\rho^{(1/2)}_{j0} &= -\sigma^{(1/2)}_{0j} = \sigma^{(1/2)}_{j0} =
-{\frac{i}{2}}\mu^{j}, \nonumber \\ \rho^{(1/2)}_{jk} &= \sigma^{(1/2)}_{jk} =
-{\frac{1}{2}}\varepsilon_{jkl} {\mu}^l,
\end{align}
$j,k,l = 1\text{-}3$. The second is
\begin{align} \label{gen1}
\rho^{(1)}_{00} &= \sigma^{(1)}_{00} = {{\bf 1}}, & \rho^{(1)}_{jk} &= \sigma^{(1)}_{jk} =
-\varepsilon_{jkl} {\cal J}^l, \nonumber
\\ \rho^{(1)}_{0j} &= -\rho^{(1)}_{j0}= -{\cal K}^j , & \sigma^{(1)}_{0j} &= - \sigma^{(1)}_{j0}
= i{\cal J}^j.
\end{align}
For a finite Lorentz transformation, ${\cal R}$ and ${\cal S}$ are
\begin{equation}\label{ACY'}
{\cal R}(\rho _{\alpha\beta}) = \exp \left ( {\frac{i}{2}} \rho _{\alpha\beta}
\zeta^{\alpha\beta} \right ), ~~ {\cal S}(\sigma_{\alpha\beta}) = \exp \left ( {\frac{i}{2}}
\sigma _{\alpha\beta} \zeta^{\alpha\beta} \right ).
\end{equation}
With the generators (\ref{gen1/2}), ${\cal R}$ and ${\cal S}$ become $R$ and $S$ of Eqs.
(\ref{RS1/2}), while the generators (\ref{gen1}) return $L$ and $M$ of Eqs. (\ref{RS1}). In the
standard $(j^{(+)},j^{(-)})$ notation for representations of the Lorentz group \cite{Brown}, the
$\rho^{(1)}_{\gamma\nu}$ generate the $(1/2,1/2)$ representation and the
$\sigma^{(1)}_{\gamma\nu}$ the $(0,0)\oplus(0,1)$ representation, while the
$\rho^{(1/2)}_{\gamma\nu}$ generate $(1/2,0) \oplus (1/2,0)$ and the $\sigma^{(1/2)}_{\gamma\nu}$
its conjugate $(0,1/2) \oplus (0,1/2)$.

For spatial rotations, ${\cal R} \equiv {\cal R}_{\rm R} = {\cal S} \equiv {\cal S}_{\rm R}$
are separately equal and unitary for either transformation (\ref{RS1/2}) or (\ref{RS1}). For
pure boosts, ${\cal R} \equiv {\cal R}_{\rm B}$ and ${\cal S} \equiv {\cal S}_{\rm B}$, we find
${\cal R}_{\rm B} = {\cal R}_{\rm B}^{\dag}$ and ${\cal S}_{\rm B} = {\cal S}_{\rm B}^{\dag}$,
again for either transformation (\ref{RS1/2}) or (\ref{RS1}) separately. However, the spin-1/2
case yields the additional relationship $S_{\rm B} = (R_{\rm B})^{-1}$, and therefore
\begin{equation}\label{SRdag}
S = (R^{\dag})^{-1} = R^{-1\dag}.
\end{equation}

Noting that the ${\mu}^{j}$ are traceless, the exponents of $R$ and $S$ in (\ref{RS1/2}) are also
traceless and thus $\text{det}{\,} R = \exp \{\text{tr}{\,} [i(\bm{\varphi} + i \bm{\zeta})
\bm{\cdot} \bm{\mu}/2]\} = +1 = \text{det}{\,} S$, as appropriate for proper Lorentz
transformations. Under {\it improper\/} transformations, ${\cal R}$ and ${\cal S}$ have the
following forms, partially found by Moses \cite{Moses1}, where $()^{\textstyle{\ast}}$ is the
complex-conjugation operator,
\begin{align}
&\text{Space inversion:} & x'^\alpha &= x_\alpha , & {\cal R}&={\cal S}=g()^{\textstyle{\ast}}
\nonumber \\ &\text{Time inversion:} & x'^\alpha &= -x_\alpha , & {\cal R}&=-{\cal
S}=g()^{\textstyle{\ast}} \nonumber \\ &\text{Spacetime inversion:} & x'^\alpha &= -x^\alpha ,
& {\cal R}&=-{\cal S}={\bf 1} .
\end{align}

\section{\label{sec:nota}INDEX NOTATION}

We employ separate index notation for four of the five vector-like representations of the
Lorentz group [$(0,0)\oplus(1,0)$ excepted], as listed in Table \ref{Table}, p.
\pageref{Table}. In this notation, we maintain the summation convention over both identical and
{\it alphabetically analogous\/} indices. For example, a term containing indices $\nu$, $n$,
and $\dot{n}$ is summed over $\nu = n = \dot{n} = 0\text{-}3 = {\it 0} \text{-} {\it 3} =
\it{\dot 0} {\text{-}} \it{\dot 3}$, or one containing $\alpha$ and $\widetilde{\alpha}$ is
summed over $\alpha = \widetilde{\alpha} = 0\text{-}3 = \widetilde{0}{\text{-}}\widetilde{3}$.
Invariant quantities are thus formed by summation over indices of the same type, e.g., $x^\beta
x_\beta$, $C^{\hat{\nu}} C_{\hat{\nu}}$, $A^a A_a$, or $Y^{\dot{a}} C_{\dot{a}}$ (Greek-tilde
indices obviously excepted). Summations of the type $x^{\alpha} A_{\dot{a}}$, $A_a
C^{\dot{a}}$, or in particular ${\partial}_{\beta} A^b = C_{\it{\dot 0}} = \Gamma$, then
generally form {\it non}-invariant quantities and thus do not ``contract out''. (Note the
apparent exceptions of $\partial_\beta G^b$ and $\partial_\alpha J^{\widetilde{\alpha}}$ in
Sec. \ref{sec:halfEM}, which are both invariant. However, $G^b$ and $J^{\widetilde{\alpha}}$
form vector-like scalars {\it only\/} with the partial derivative operator, not under
contraction with an arbitrary vector.) Finally, in an equation with mixed indices on separate
terms, such as $X^{\widetilde{\mu}} = A^m + B^{\mu}$, alphabetically analogous indices like
$\widetilde{\mu}$, $m$, and $\mu$ will be understood to take the same values $\widetilde{\mu} =
m = \mu = \widetilde{0}{\text{-}}\widetilde{3} = {\it 0}\text{-}{\it 3} = 0\text{-}3$. We
assume the following correspondence between Latin and Greek indices: $a \alpha$, $b \beta$, $c
\chi$, $d \delta$, $e \epsilon$, $f \phi$, $g \gamma$, $h \eta$, $i \iota$, $k \kappa$, $l
\lambda$, $m \mu$, $n \nu$, $p \pi$, $r \rho$, $s \sigma$, $t \tau$, $u \upsilon$, $w \omega$,
$x \xi$, $y \psi$, $z \zeta$.

\section{\label{sec:basis}SPIN-1/2 AND SPIN-1 MINKOWSKI-BASIS EIGENVECTORS}

We present here the Minkowski basis eigenvectors for both spin-1 and spin-1/2 fields (compare Ref.
\onlinecite{Moses2}). In the familiar vector case, the spin matrices associated with $L$ and $M$
are ${\cal J}_x$, ${\cal J}_y$, ${\cal J}_z$, and $\bm{{\cal J}} \equiv {\bf \hat{x}}{\cal J}_x +
{\bf \hat{y}}{\cal J}_y + {\bf \hat{z}}{\cal J}_z$. ${\cal J}_z$ and ${\bm{{\cal J}}}^2$ have
simultaneous spin eigenvalues $s = 0,1$ and $m_s = -1,0,+1$, with eigenfunctions $^\epsilon {\rm
X}$ satisfying ${\cal J}_z {\,}^{\epsilon} {\rm X} = m_s {\,}^{\epsilon} {\rm X}$ and $\bm{{\cal
J}}^{2} {\,}^{\epsilon} {\rm X} = s(s+1) {\,}^{\epsilon} {\rm X}$ where the left index $\epsilon$
numbers the eigenfunctions. Defining raising and lowering operators ${\cal J}_{\pm} \equiv {\cal
J}_{x} {\pm} i {\cal J}_{y}$, the $^{\epsilon}{\rm X}$ resolve into four orthogonal basis vectors
$^{\epsilon}{d}$, $\epsilon = 0\text{-}3$: $^{0}d = \text{col}(1,0,0,0)$, $^{1}d =
\text{col}{\frac{1}{\sqrt{2}}}(0,1,-i,0)$, $^{2}d = \text{col}(0,0,0,1)$, and $^{3}d =
\text{col}{\frac{1}{\sqrt{2}}}(0,-1,-i,0)$. $^0 d$ represents $s = m_s = 0$ while $^{1,2,3} d$ are
$s = 1$ with $m_s = -1, 0, +1$ respectively.

In the analogous spin-1/2 formalism, the spin matrices are ${\rm s}_x \equiv \mu^{1}/2$, ${\rm
s}_y \equiv \mu^{2}/2$, ${\rm s}_z \equiv \mu^{3}/2$, and \vspace{0mm}
\begin{equation}
\bm{{\rm s}} \equiv {\frac{1}{2}}({\bf{\hat{x}}} \mu^{1} + {\bf{\hat{y}}}\mu^{2} +
{\bf{\hat{z}}}\mu^{3}) = {\bm{\mu}},
\end{equation}
where $[{\rm s}^j , {\rm s}^k] = i \varepsilon_{jkl} {\rm s}^l$. ${\rm s}_z$ and $\bm{{\rm s}}^2$
have simultaneous eigenfunctions $^{\epsilon}{\rm Z}$ satisfying
\begin{equation}
{\rm s}_z {\,}^{\epsilon} {\rm Z} = m_s {\,}^{\epsilon} {\rm Z} = \pm {\frac{1}{2}}{\,}^{\epsilon}
{\rm Z} ~~~\text{and}~~~ {\bm{{\rm s}}}^{2} {\,}^{\epsilon} {\rm Z} = s ( s + 1 ) {\,}^{\epsilon}
{\rm Z} = {\frac{3}{4}}{\,}^{\epsilon} {\rm Z}
\end{equation}
so that $s = 1/2$ and ${m}_{s} = \pm 1/2$. Defining ${\rm s}_{\pm} \equiv {\rm s}_x \pm i {\rm
s}_y = (\mu^{1}{\pm} i \mu^{2})/2$, the $^{\epsilon}{\rm Z}$ resolve into four orthogonal basis
spinors $^{\epsilon}{b\/}$, $\epsilon = 0\text{-}3$, grouping into two pairs, even $^{0} b =
(^0 d + {\,}^2 d)/{\sqrt{2}} = {\downarrow}_{even}$, $^{2} b = {\,}^{3} d = {\uparrow}_{even}$,
and odd $^{1} b = -{\,}^{1} d = {\downarrow}_{odd}$, $^{3} b = (^0 d - {\,}^2 d)/{\sqrt{2}} =
{\uparrow}_{odd}$ satisfying $^{\gamma} b ^{\dag } {\,}^{\epsilon} b = \delta _{\gamma
\epsilon}$, $\sum_{\epsilon = 0}^{3} {\,} ^{\epsilon} {b_a}^{\textstyle \ast} {\,}^{\epsilon}
{b} _{b} = \delta _{a b}$, and ${\,}^{\epsilon} b ^{a} {\,}^{\epsilon} {b} _{a} = 0$ (no sum
over $\epsilon$). The $^{\epsilon} d$ and the $^{\epsilon} b$ are related by a real linear
transformation, neither are intrinsically spin 1 or spin 1/2.


\begin{thebibliography}{}
\bibitem{clear} We distinguish here between instances where Maxwell variables transform with
1/2-angle rules and those where they transform as a varying part of a larger spinor field while
retaining their usual spin-1 invariance properties. We refer only to the former as a spin-1/2
Maxwell field. The latter occurs, for example, with the generalized irreducible Maxwell equations
of Sachs and Schwebel \cite{Sachs1}, or Leiter's Dirac-spinor Maxwell equations \cite{Leit}. In
these cases, the spinor dependence on Maxwell variables necessarily changes from frame to frame,
allowing the Maxwell field and current to maintain their usual invariance relations despite
transforming within a governing spinor field. Neither is the Maxwell-Dirac--equivalent field of
Campolattaro, Rodrigues, and Vaz \cite{Camp,RV} a spin-1/2 Maxwell field. In this case the authors
describe a nonlinear spinor equation equivalent to Maxwell's equations. But the Maxwell field
remains a vector and tensor composition of the spinor field, not itself a spinor field, and thus
its standard transformation properties are again unchanged.
\bibitem{reps} Excepting the 4-scalar representation, these are the only representations of
the Lorentz group that form invariants with the Minkowski metric {\it g\/} (or \d{\it g}), as
may be shown by observation of the remaining four-component representations.
\bibitem{hispin} Higher rank versions of four-spinors can similarly be formed as in the
two-component case
\begin{equation*}\label{xiabc}
{{Z'} } _{ b\cdots \dot{d} }^{a\cdots \dot{c} } = {{R} ^{a} }_{e} \cdots {{R} ^{\dot{c} } }
_{\dot{g} } \cdots {R _{b} } ^{f} \cdots {R _{\dot{d} } } ^{\dot{h} } \cdots Z_{f \cdots
\dot{h} }^{ e\cdots \dot{g} } .
\end{equation*}
\bibitem{8x8} Similar eight-component matrix equations are in \onlinecite{Sim2,8comp},
and Borgardt \onlinecite{Oppy}.
\bibitem{light} Among the many, Sallhofer describing his electromagnetic model of
the Dirac field: ``It depicts the hydrogen field as a light wave impinging ... on a point, being
reflected by itself at that point ... The incoming and outgoing waves form a standing wave field
... [and the] concentration of energy at the center is what we perceive as the hydrogen atom''
\cite{Sal1}.
\end{thebibliography}
\end{document}